\documentclass[pra,reprint,twocolumn,10pt,showpacs,superscriptaddress,nofootinbib]{revtex4-1}
\usepackage{amsmath}
\usepackage{amsthm}
\usepackage{graphicx}
\usepackage{times}
\usepackage{amssymb}
\usepackage{hyperref}
\hypersetup{
colorlinks=true,
linkcolor=red,
citecolor=red,}
\usepackage{textcomp}
\usepackage{xcolor}
\usepackage{enumerate}
\usepackage{multirow}
\usepackage{tabularx}
\usepackage{dcolumn}
\usepackage[mathscr]{euscript}
\usepackage{mathtools}
\usepackage{needspace}
\usepackage[mathscr]{euscript}

\newcommand{\etal}{\emph{et al}}
\newcommand{\PRA}[4]{#3 #4 \emph{Phys. Rev. A} \textbf{#1} #2}

\newcommand{\PRL}[4]{#3 #4 \emph{Phys. Rev. Lett.} \textbf{#1} #2}
\newcommand{\RMP}[4]{#3 #4 \emph{Rev. Mod. Phys.} \textbf{#1} #2}

\newcommand{\Science}[4]{#3 #4 \emph{Science} \textbf{#1} #2}

\newcommand{\Nature}[4]{#3 #4 \emph{Nature} \textbf{#1} #2}
\newcommand{\NatPhys}[4]{#3 #4 \emph{Nat. Phys.} \textbf{#1} #2}
\newcommand{\NatPhot}[4]{#3 #4 \emph{Nature Photon.} \textbf{#1} #2}

\newcommand{\PhysRep}[4]{#3 #4 \emph{Phys. Rep.} \textbf{#1} #2}
\newcommand{\Optica}[4]{#3 #4 \emph{Optica} \textbf{#1} #2}
\newcommand{\TC}[4]{#3 #4 \emph{Theory of Computing} \textbf{#1} #2}
\newcommand{\MC}[4]{#3 #4 \emph{Math. Comput.} \textbf{#1} #2}
\newcommand{\SR}[4]{#3 #4 \emph{Sci. Rep.} \textbf{#1} #2}
\newcommand{\JOB}[4]{#3 #4 \emph{J. Opt. B: Quantum Semiclassical Opt.} \textbf{#1} #2}
\newcommand{\SA}[4]{#3 #4 \emph{Sci. Adv.} \textbf{#1} #2}
\newcommand{\OL}[4]{#3 #4 \emph{Opt. Lett.} \textbf{#1} #2}
\newcommand{\OEx}[4]{#3 #4 \emph{Opt. Exp.} \textbf{#1} #2}

\begin{document}

\title{Arbitrary Multi-Qubit Generation}
\author{F. Shahandeh}
\email{Email correspondence: f.shahandeh@uq.edu.au}
\author{A. P. Lund}
\author{T. C. Ralph}
\affiliation{Centre for Quantum Computation and Communication Technology, School of Mathematics and Physics, University of Queensland, St Lucia, Queensland 4072, Australia}
\author{M. R. Vanner}
\email{Email correspondence: michael.vanner@physics.ox.ac.uk}
\affiliation{Centre for Quantum Computation and Communication Technology, School of Mathematics and Physics, University of Queensland, St Lucia, Queensland 4072, Australia}
\affiliation{Clarendon Laboratory, Department of Physics, University of Oxford, OX1 3PU, United Kingdom}
\date{\today} 

\begin{abstract}
We propose and analyse a scheme for single-rail-encoded arbitrary multi-qubit quantum-state generation to provide a versatile tool for quantum optics and quantum information applications. 
Our scheme can be realized, for small numbers of qubits, with current technologies using single photon inputs, passive linear optics, and heralding measurements.
The particular examples of two- and three-qubit cluster states are studied in detail.
We show that such states can be prepared with a high probability of success.
Our analysis quantifies the effects of experimentally relevant imperfections and inefficiencies.
The general case of arbitrary $N$-qubit preparation is discussed and some interesting connections to the boson sampling problem are given.
\end{abstract}

% \pacs{03.67.Mn, 42.50.Dv}

\maketitle

\section{Introduction}

An active field of research in quantum optics and quantum information is the development of techniques for producing arbitrary quantum states of different physical systems.
This is by virtue of the broad range of applications including quantum computation and communication~\cite{Nielsen,Kimble2008}, quantum simulation~\cite{Guzik2012}, and quantum metrology~\cite{Giovannetti2011}, that each need different specific quantum states as a resource.

Light is a key quantum system which can be interfaced with a variety of other individual quantum systems~\cite{Leibfried2003, Hammerer2010, Doherty2013}.
There are particular states, e.g., W-states~\cite{Glockl2003}, GHZ-states~\cite{Patel2016}, and cluster states~\cite{Browne2005,Lu2007}, which are of great importance for many protocols, and it is anticipated that the exploration of other multimode states will also be promising. 
Thus, having a single easily reconfigurable device which prepares arbitrary multimode quantum states would provide considerable versatility for numerous applications~\cite{Carolan2015,Schaeff2015}.
Towards this goal, a number of theoretical and experimental studies have been performed concerning the preparation of arbitrary single-mode~\cite{Dakna1999,Yukawa2013,Brannan2014} as well as arbitrary multimode quantum states~\cite{VanMeter2007,Andersen2015}.
It is usually the case that universal schemes possess a high degree of complexity due to their extremely large number of degrees of freedom, even for a small number of modes, which makes them impractical without some restrictions.
A feasible approach to this problem can be taken if we restrict the dimensionality of each output mode to two.
In other words, we restrict ourselves to {\it arbitrary multi-qubit state generation}.
Regarding the cost of the scheme, we will be utilising passive linear transformations; avoiding any active elements, which in turn means to only use beam splitters and phase shifters, and we allow measurements on some ancillary systems to introduce the required nonlinearity.

We would also like to encode quantum bits in the absence or presence of photons, the so-called single-rail encoding~\cite{Lund2002}.
As well as requiring fewer resources, such states are efficient under temporal multiplexing~\cite{Simon2007} and suitable for interactions with matter quantum systems.
The key goal is thus a device which produces arbitrary single-rail multi-qubit resource states for a variety of quantum tasks.
Importantly, we want to prepare the target state by heralding via measurements of the ancillae and without postselection.
To the best of authors' knowledge, our scheme is the first of this kind which provides all of the properties mentioned above.

In the present contribution, we introduce and analyse a universal scheme for generation of arbitrary multi-qubit quantum states in a single-rail encoding using passive linear optics and heralding measurements.
Our scheme makes use of $N$ single-photon input states.
We extract part of their amplitude using beam splitters and inject it into a unitary network of size $M(N)$.
The other ports of the unitary will receive $K$ ancillary single-photon inputs and $M-N-K$ ancillary vacuum inputs, while there will be $L$ projections onto single-qubits and $M-L$ vacuum measurements at the output.
We show our scheme is universal and give particular examples of two- and three-qubit cluster states as target states, a class of states known to be very hard to prepare in single-rail encoding.
We show that our scheme delivers a high probability of success, for low numbers of photons, even compared to non-universal schemes where there exists a fairly comparable strategy.
We also analyse the effects of loss and imperfections.
Our scheme is experimentally feasible for a small number of photons using current technology.
Moreover, we study the general case of $N$-qubit generation and give interesting connections to the boson sampling problem~\cite{AA}.
The method presented here is equally applicable to other bosonic systems, e.g. spin ensembles~\cite{McConnell2015} and optomechanics~\cite{Vanner2013} by making use of light-matter beam splitter interactions and photonic ancillae.

%==============================================================================
%==============================================================================
\section{Arbitrary two-qubit state preparation}

In this section we design a linear-optics scheme to generate arbitrary two-qubit target states,
\begin{equation}
\label{TwoTarg}
	|\psi_{\rm tar}\rangle = c_{00}|00\rangle + c_{01}|01\rangle + c_{10}|10\rangle + c_{11}|11\rangle,
\end{equation}
with $\{c_{ij}\}$ ($i,j=0,1$) being complex numbers.
Such a state can be for example shared between two distant parties, Alice and Bob.

\begin{figure}[h]
  \includegraphics[width=7cm]{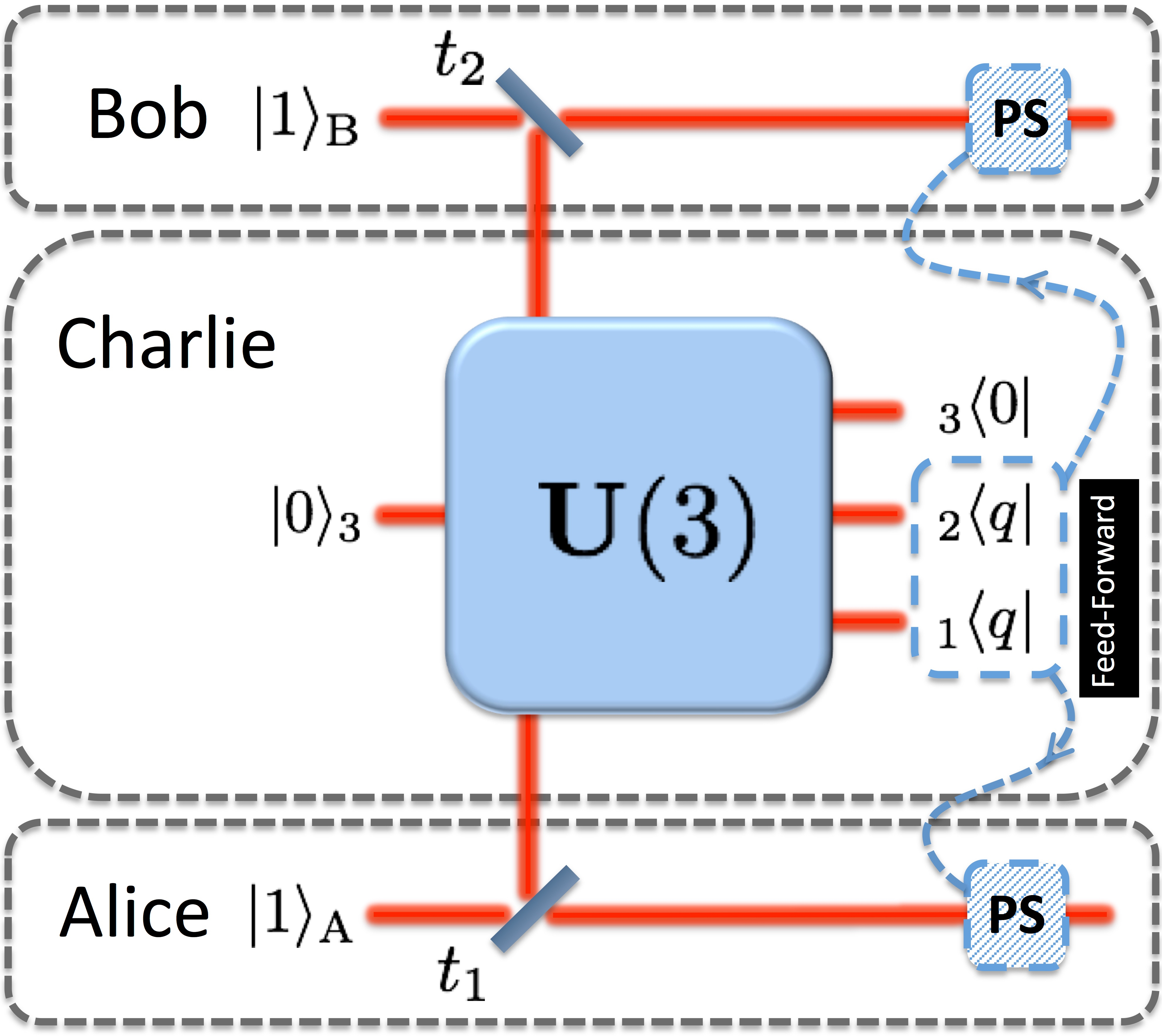}
  \caption{Scheme for two-qubit state generation.
  Alice and Bob split their single-photons and send it to Charlie.
  He makes an entangling measurement and remotely prepares arbitrary joint states for Alice and Bob.
  The blue dashed boxes refer to the optional feed-forward correction of the outputs depending on the single-qubit measurements outcomes.
  }\label{TwoQBSch}
\end{figure}

A major problem in any such a device then would be the suppression of the two-photon terms $|02\rangle$ and $|20\rangle$ due to Hong-Ou-Mandel effect.
It has been shown that without relying on measurements one cannot avoid these bunching effects~\cite{Wu}.
A simple way to circumvent this issue is by preventing the photons from exchanging between two principal modes from input to output, rather than using Fock basis truncations~\cite{Pegg}; see Fig.~\ref{TwoQBSch}.
We will show that this technique despite its simplicity is very effective.

Accordingly, we split the circuit into two stages: (i) an entangling projector
-- this could be carried out by a third party, say Charlie, as a central station who sends signals to Alice and Bob on the successful events, and (ii) a quantum {\it pick-off}; this could be done locally with the help of Alice and Bob (c.f. Fig.~\ref{TwoQBSch}).
In the following we give a detailed description of both stages.

%============================================
\subsection{Entangling measurement}

Firstly, let us discuss what we mean by entangling measurement in our scheme.
We note that to be able to produce the state in~\eqref{TwoTarg}, there is a need for independent control of four coefficients $\{c_{ij}\}$ ($i,j=0,1$).
For the moment, we do not take into account the reduction of degrees of freedom by normalization constraint, because it should be possible to equate any of the coefficients with zero.
This demands Charlie to realize projectors of the form
\begin{align}
\label{TwoProj}
 \langle\varphi|=\langle 00|d_{00} & + \langle 01|d_{01} + \langle 10|d_{10} + \langle 11|d_{11} \nonumber\\
 & \quad + \langle 02|d_{02} + \langle 20|d_{20}, %\in \mathscr{H}_{2}^{\ast\otimes 2},
\end{align}
such that the first four complex coefficients, $\{d_{ij}\}$ ($i,j=0,1$), are arbitrarily controllable.
Notice that the need for two-photon components arises from the use of linear optics as previously commented.
As we will show, a $\hat{U}(3)$ unitary operation with an ancillary vacuum measurement and two arbitrary single-qubit projectors is the smallest unitary circuit having the capability to produce $\langle\varphi|$.
Using a retrodictive approach, we now calculate the projector in Fig.~\ref{TwoQBSch} just before the unitary.
This is given by
\begin{align}
\label{retroOrig}
	_{123}\langle\varphi|&={_{123}\langle} q_1q_20|\hat{U}(3)\nonumber\\
    &={_{123}\langle 000|}(\alpha_1 \mathbb{I}_1+\beta_1 a_1){\otimes}(\alpha_2 \mathbb{I}_2+\beta_2 a_2)\hat{U}(3),
\end{align}
in which $_i\langle q_i|={_i\langle 0|}(\alpha_i \mathbb{I}_i+\beta_i a_i)$ ($i=1,2$) with $|\alpha_{1,2}|^2+|\beta_{1,2}|^2=1$.
Using the definition $\hat{U}^\dag(3) a_i \hat{U}(3)=\sum^3_{j=1} u_{ij}a_j$ ($i=1,2,3$), we now switch to the Heisenberg picture via the isomorphic matrix representation,
\begin{equation} 
\label{U3}
	\mathbf{U}(3)=
	\begin{bmatrix}
    \mathbf{S}(2) & \begin{matrix} u_{13} \\ u_{23} \end{matrix} \\
    \begin{matrix} u_{31} & u_{32} \end{matrix} & 
    u_{33}
\end{bmatrix}
\end{equation}
in which
\begin{equation}
\label{S2}
	\mathbf{S}(2)=
	\begin{bmatrix}
	u_{11} & u_{12}\\ 
	u_{21} & u_{22}
	\end{bmatrix}.
\end{equation}
We call the submatrix $\mathbf{S}(2)$ the {\it active} submatrix.
Rearranging the resulting components in~\eqref{retroOrig} and projecting onto the ancillary vacuum input, $| 0 \rangle_3$, gives
\begin{align}
\label{phi12}
	_{12}\langle\varphi| & = {_{12}\langle} 00| d_{00} +{_{12}\langle} 01| d_{01} + {_{12}\langle} 10|d_{10} + {_{12}\langle}11|d_{11} \nonumber\\
	& \qquad \qquad \qquad \qquad +{_{12}\langle}02| d_{02} +{_{12}\langle}20| d_{20},
\end{align}
in which
\begin{align}
 & d_{00} = \alpha_1\alpha_2, \label{d00}\\
 \begin{split}
 & d_{01} = \alpha_1\beta_2 \mathcal{P}_{2;2}+\alpha_2\beta_1 \mathcal{P}_{1;2}, \\
 & d_{10} = \alpha_1\beta_2 \mathcal{P}_{2;1}+\alpha_2\beta_1 \mathcal{P}_{1;1}, \\
 & d_{11} = \beta_2\beta_1 \mathcal{P}_{12;12},\label{3Eqs}
 \end{split}\\
 \begin{split}
 & d_{02} = \beta_2\beta_1\mathcal{P}_{1;2}\mathcal{P}_{2;2},\\
 & d_{20} = \beta_2\beta_1 \mathcal{P}_{1;1}\mathcal{P}_{2;1}.
 \end{split}\label{2Extra}
\end{align}
Here\footnote{These equations result from following photons paths, e.g. in the first relation in Eq.~\eqref{3Eqs}, there are two contributions to $d_{01}$, that is the case when a single photon is input into port two and no photons into port one: (i) either a photon is detected at the output mode two with probability amplitude of $\beta_2$ coming from mode two with a transition probability amplitude of $\mathcal{P}_{2;2}$, (ii) or a photon is detected at the output mode one with probability amplitude of $\beta_1$ coming from mode two with a transition probability amplitude of $\mathcal{P}_{1;2}$. Also $\alpha_{1,2}$ represent the probability amplitudes of getting no photons in the other port.}, $\mathcal{P}_{ij\cdots p;kl\cdots q}$ represents the \textit{permanent} of the active submatrix, which are obtained from elements on the intersection of rows $i,j,\dots,p$ and columns $k,l,\dots,q$.
Note that, permanents have cyclic symmetry, that is $\mathcal{P}_{ij\cdots p;kl\cdots q}=\mathcal{P}_{\sigma(i)\sigma(j)\cdots\sigma(p);\sigma(k)\sigma(l)\cdots\sigma(q)}$ for any permutation $\sigma$ of the index set, and $\mathcal{P}_{i;j}=u_{ij}$ trivially.

%============================================
\subsection{Pick-off}

At the second stage, we need to chop off the extra dimensions of our entangling measurement.
For this, Alice and Bob split their single-photons
% we split the principal modes injected by two single-photon inputs 
using two variable beamsplitters, characterized by transmissivity-reflectivity ratios $t_1:r_1$ and $t_2:r_2$ (c.f. Fig~\ref{TwoQBSch}).
That is, $|01\rangle_{1\rm{A}}\otimes|01\rangle_{2\rm{B}}\rightarrow|b_1\rangle_{1\rm{A}}\otimes |b_2\rangle_{2\rm{B}}=(t_1|01\rangle_{1\rm{A}} + r_1|10\rangle_{1\rm{A}})\otimes(t_2|01\rangle_{2\rm{B}} + r_2|10\rangle_{2\rm{B}})$ upon beamsplitter transformation $\hat{B}(a^\dag_1, a^\dag_{\rm A})\hat{B}^\dag = (t_1a^\dag_1 -r_1a^\dag_{\rm A},r_1a^\dag_1 +t_1a^\dag_{\rm A})$.
Now, Charlie applies his entangling projector on two branches he receives from Alice and Bob, and thus the un-normalized output is\footnote{This equation can also be readily interpreted. For instance, the first term can be thought of as the product of probability amplitudes of: (i) both input photons are reflected from input beamsplitters, and thus directed toward entangling device, and (ii) triggering the measurement device upon receiving two single photons.}
\begin{align}
\label{Out}
	|\psi_{\rm out}^{11}\rangle & = {_{12}\langle\varphi|}(|b_1\rangle_{1\rm{A}}\otimes |b_2\rangle_{2\rm{B}})\nonumber\\
	& = r_1r_2d_{11}|00\rangle_{\rm{A}\rm{B}} + r_1t_2d_{10}|01\rangle_{\rm{A}\rm{B}} \nonumber\\
	& \qquad + t_1r_2d_{01}|10\rangle_{\rm{A}\rm{B}} + t_1t_2d_{00}|11\rangle_{\rm{A}\rm{B}},
\end{align}
in which $|t_{1,2}|^2+|r_{1,2}|^2=1$ and $\{d_{ij}\}$ ($i,j=0,1$) are given by Eqs.~\eqref{d00} and~\eqref{3Eqs}.
For future reference, we have used the superscript $11$ to indicate that the input was $|11\rangle_{\rm AB}$.
The probability of success for getting the desired target state is then given by the trace of output state, $\mathrm{Pr}_{\rm tar}={\rm Tr}|\psi_{\rm out}^{11}\rangle\langle\psi_{\rm out}^{11}|$.
By comparing Eq.~\eqref{Out} with Eq.~\eqref{TwoTarg}, we can write down a set of nonlinear equations to be solved for all parameters, such that the probability of success is optimized:
\begin{align}
\label{Set1}
\begin{split}
& \frac{r_1r_2\beta_1\beta_2\mathcal{P}_{12;12}}{p}-c_{00}=0,\\
& \frac{r_2t_1(\alpha_1\beta_2\mathcal{P}_{2;2}+\alpha_2\beta_1\mathcal{P}_{1;2})}{p}-c_{10}=0\\
& \frac{r_1t_2(\alpha_1\beta_2\mathcal{P}_{2;1}+\alpha_2\beta_1\mathcal{P}_{1;1})}{p}-c_{01}=0\\
& \frac{t_1t_2\alpha_1\alpha_2}{p}-c_{11}=0\\
& {\rm Pr_{tar}}=|p|^2 =|r_1r_2d_{11}|^2 + |r_1t_2d_{10}|^2 \\
& \qquad\qquad\qquad\qquad + |t_1r_2d_{01}|^2 + |t_1t_2d_{00}|^2.
\end{split}
\end{align}
There are also constraints to be satisfied by the $\mathbf{U}(3)$ entries: (i) each row  must be normalized, and (ii) each row must be orthogonal to all other rows.
Let us show that a $\mathbf{U}(3)$ matrix is the smallest unitary matrix for which the set~\eqref{Set1} possesses a solution under the aforementioned constraints.
First, we note that $c_{11}$ is uniquely determined by $d_{00}$ through $\alpha_1$ and $\alpha_2$, independent of the unitary.
Therefore, we need to control three amplitudes and only three phases using the unitary.
Second, notice that the matrix $\mathbf{U}(3)$ in Eq.~\eqref{U3} has six free parameters, three amplitudes and three phases.
Also, one should note that one of the phases is global which could determine the relative phase between $|11\rangle$ and the other components.
However, this can be also absorbed into the phase of $\alpha_{1,2}$.
As a result, only two phases are sufficient.
Nevertheless, due to the fact that three free amplitudes are required, the smallest matrix would be that of a $\mathbf{U}(3)$.
In Appendix~\ref{AppA}, we have shown that the set of equations~\eqref{Set1} indeed possesses a solution when there are three complex variables available.
In the next sections we will first describe how it is possible to perform single-qubit projections, and then we demonstrate our solution strategy to these equations through an example.

%============================================
\subsection{Single-qubit projectors}\label{SQProj}

In the scheme already discussed, two arbitrary single-qubit projectors at the output of the unitary network play crucial roles.
For instance, their coefficient directly determine the coefficient of $|11\rangle$ component; see e.g. Eqs.~\eqref{d00} and~\eqref{Out}.
Here we describe how these measurements can be realized and discuss the associated heralding probabilities.

\begin{figure}[h]
  \includegraphics[width=4cm]{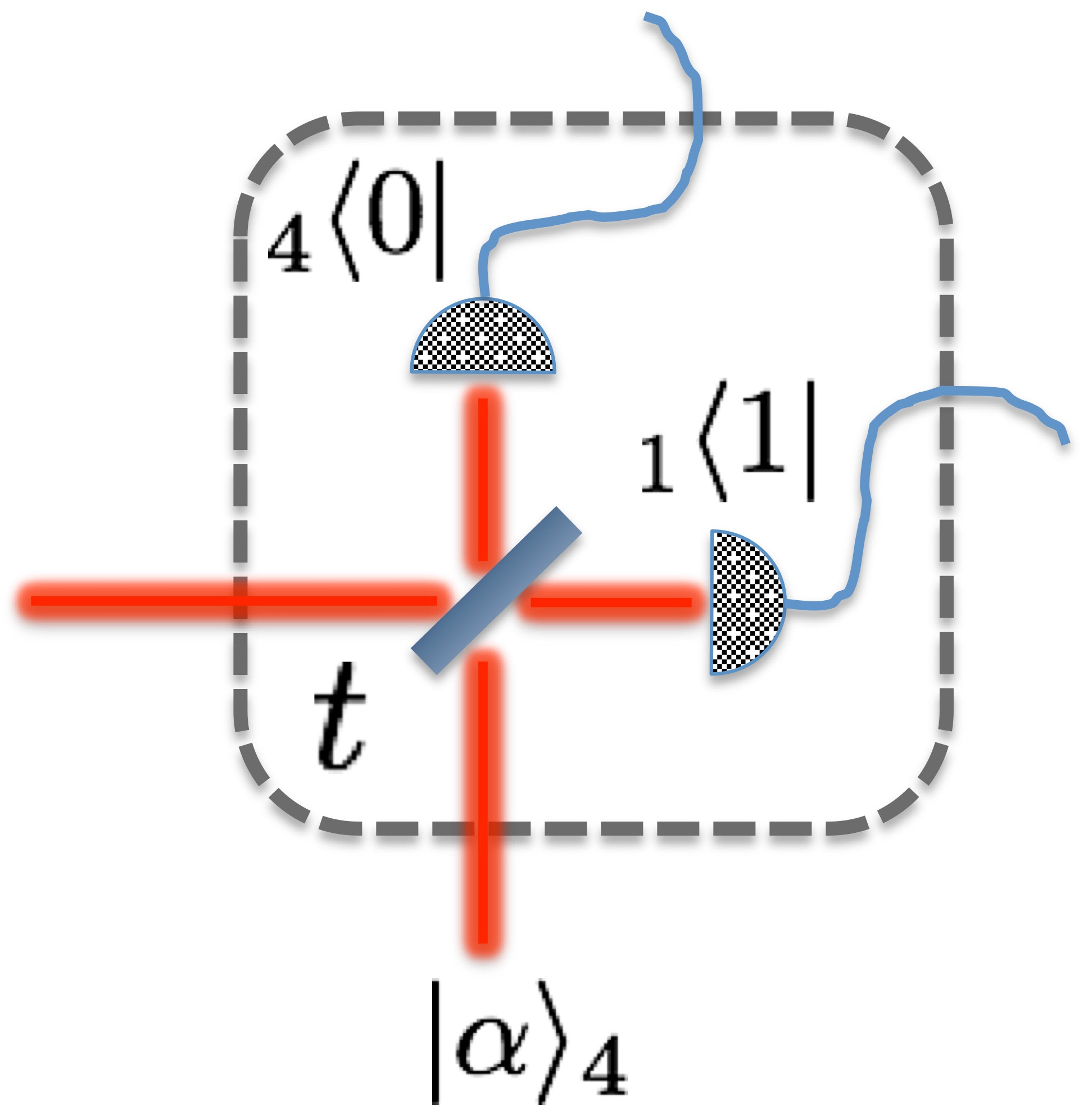}
  \caption{The scheme for probabilistic projection onto a single-qubit basis using a beam splitter, a coherent state input, and two single-photon counters. 
  }\label{SQProjCoh}
\end{figure}

First, let us propose a simple way of implementing our arbitrary single-qubit projector.
In this proposal, a coherent state $|\alpha\rangle_{4}$ hits a beamsplitter of trasmissivity-reflectivity $t:r$; see Fig.~\ref{SQProjCoh}.
For simplicity, we are assuming both trasmissivity and reflectivity take on real values.
The output is then measured by two single-photon counters.
The beamsplitter transformation is assumed to be $\hat{B}(a^\dag_{1}, a^\dag_{4})\hat{B}^\dag = (ta^\dag_{1} -ra^\dag_{4},ra^\dag_{1} +ta^\dag_{4})$ with $t^2+r^2=1$, thus, the resulting projector can be calculated as
\begin{align}
\label{10C}
   _{1}\langle q|= {_{14}}\langle 10|\hat B|\alpha\rangle_4=
    e^{-\frac{|\alpha|^2}{2}}({_1}\langle 0|r\alpha + {_1}\langle 1|t).
\end{align}
Choosing $\alpha=1$ can generate any single-qubit measurement with probability of $e^{-1}$, while the phases can be implemented via some phase shifts on beamsplitter.

The $e^{-1}$ overhead from this scheme is entirely due to the exponential prefactor which originates from the use of a coherent state input.  
This overhead could be eliminated if the state were confined to the lowest two Fock basis states.  
Therefore, inspired by the work of Ralph \etal~\cite{Lund} which proposes the use of real-time adaptive detection of phase from~\cite{Wiseman98} to {\it deterministically} prepare superposition states of zero and one photon, we propose the following.
Assuming that one can deterministically prepare arbitrary single-qubit states via adaptive phase measurement,
the coherent state $|\alpha\rangle_{4}$ is replaced by a single-qubit state $\beta|0\rangle_{4} + \alpha|1\rangle_{4}$ and the beamsplitter is being fixed to be balanced.
At the output, there are three possible events containing different combinations of zero and one clicks: $(1,0)$, $(0,1)$, and $F$ (the failure event).
Considering the $(1,0)$ event, using a retrodictive approach, we find
\begin{align}
\label{10E}
   _{1}\langle q|= {_{14}}\langle 10|\hat B(\beta |0\rangle_4 + \alpha |1\rangle_4)=
    \frac{1}{\sqrt 2}({_1}\langle 0|\alpha + {_1}\langle 1|\beta),
\end{align}
This is half of the desired event, in the sense that if we inject the correct state $|q\rangle_1$ into the device, only half of the times it will pop-up the occurrence of the event.
There are, however, two other possible events:
$F$, the total failure event, and $(0,1)$.
The latter is of particular interest as we will see shortly.
If we get the event $(0,1)$ instead of $(1,0)$, we have
\begin{align}
\label{01E}
    _{1}\langle \tilde{q}|= {_{14}}\langle 01|\hat B(\beta |0\rangle_4 + \alpha |1\rangle_4)=
    \frac{1}{\sqrt 2}({_1}\langle 0|\alpha - {_1}\langle 1|\beta).
\end{align}
In the following, we show that one could correct the output state simply by feed-forwarding two extra classical bits to the principal output modes.
Notably, in general, the two positive operator-valued-measure (POVM) elements $\Pi^{(1)}=|q\rangle_{1}\langle q|$ and $\Pi^{(2)}=|\tilde q\rangle_{1}\langle \tilde q|$ are not orthogonal.
The failure event is also given as $F\leftrightarrow\Pi^{(0)}=\mathbb{I}-\Pi^{(1)}-\Pi^{(2)}$.

Note that the event space of the clicks for our two-qubit scheme is nine fold: $\{\Pi^{(i,j)}\equiv\Pi^{(i)}\otimes\Pi^{(j)}\}$ ($i,j=0,1,2$), and only one of them is the desired click, namely $\Pi^{(1,1)}$, corresponding to two $(0,1)$ events of the form~\eqref{10E} and resulting the projection $_{12}\langle\varphi|/2$ of Eq.~\eqref{phi12}.
Importantly, the factor of $1/2$ causes the overall probability of success to reduce by a factor of $4$ with respect to the ideal case.

Now, suppose that we get the {\it invert} event $\Pi^{(2,2)}$.
In view of Eq.~\eqref{01E}, this will result in two phase shifts of $\pi$ in $\beta_{1,2}$ and thus,
according to Eqs.~\eqref{d00},~\eqref{3Eqs}, and~\eqref{Out}, nothing is changed except the two coefficients $d_{01}$ and $d_{10}$ which pick up an extra phase of $\pi$.
Evidently, the invert output occurs with exactly the same probability of success as the desired one.
Moreover, we may correct the invert output by feed-forwarding the outcome to apply two local phase shifts to the principal modes; c.f. Eq.~\eqref{Out}.
Such phase shifts will change the sign of the coefficients of $|01\rangle$ and $|10\rangle$ terms leaving the $|11\rangle$ component intact.

We may ask how much does the correction improve the overall probability of success?
To answer, we note the probability of success of invert event to be the same as desired event.
This implies that taking into account the possibility of feed-forward correction doubles the overall probability of success, and decreases its reduction due to nondeterministic measurements to a factor of $2$ with respect to the ideal case.

%============================================
\subsection{Example: A cluster state}

\begin{figure}[h]
  \includegraphics[width=8cm]{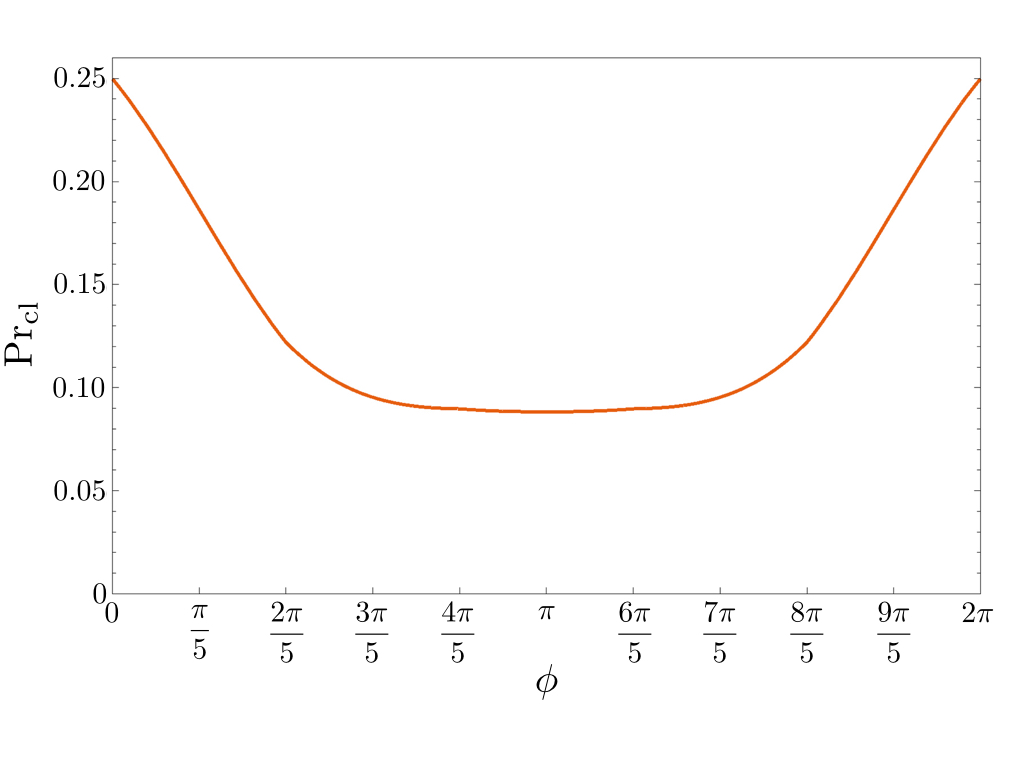}
  \caption{Plot of optimal success probability versus relative phase, $\phi$, of the generalized two-qubit cluster state given in Eq.~\eqref{TwoCluster} for $\chi=0$.
  }\label{TwoQBClustPhi}
\end{figure}

Let us consider the example of preparing the state,
\begin{equation}
\label{TwoCluster}
	|\psi_{\rm cl}\rangle = \frac{e^{i\chi}}{2}(|00\rangle + |01\rangle + |10\rangle + e^{i\phi}|11\rangle),
\end{equation}
which gives the usual two-qubit cluster state when $\phi=\pi$.
We consider the output state~\eqref{Out} and without loss of generality assume $\alpha_{1,2}\rightarrow\alpha_{1,2}e^{i\theta}$
\footnote{Any phase difference between $\alpha_{1,2}$ can be absorbed into the unitary elements, as they can be represented by some phase shifters in front of the unitary circuit.} with $\alpha_{1,2}\in\mathbb{R}$ 
and that $\beta_{1,2}=\sqrt{1-\alpha_{1,2}^2}\in\mathbb{R}$.
We also take $t_1$ and $t_2$ to be real.
Similar to Eq.~\eqref{Set1}, these lead to the following set of nonlinear equations
\begin{equation}
\label{setU}
\begin{split}
& \theta = \phi - \zeta,\\ 
& \left( \frac{\sqrt{1-\alpha_2^2}}{\alpha_2} \mathcal{P}_{2;2} + \frac{\sqrt{1-\alpha_1^2}}{\alpha_1} \mathcal{P}_{1;2}\right) \frac{\sqrt{1-t_2^2}}{t_2} e^{i\zeta} = 1,\\ 
& \left(\frac{\sqrt{1-\alpha_2^2}}{\alpha_2}\mathcal{P}_{2;1} + \frac{\sqrt{1-\alpha_1^2}}{\alpha_1}\mathcal{P}_{1;1}\right)\frac{\sqrt{1-t_1^2}}{t_1}  e^{i\zeta} = 1,\\
& \frac{\sqrt{(1-\alpha_1^2)(1-\alpha_2^2)(1-t_1^2)(1-t_2^2)}}{\alpha_1\alpha_2t_1t_2} \mathcal{P}_{12;12} e^{i(2\zeta - \phi)}= 1,
\end{split}
\end{equation}
in which $\zeta \equiv \theta - \chi$, and it is constrained to
\begin{equation}
\label{setConst}
\begin{split}
& |\mathcal{P}_{1;1}|^2+|\mathcal{P}_{1;2}|^2+|\mathcal{P}_{1;3}|^2=1,\\
& |\mathcal{P}_{2;1}|^2+|\mathcal{P}_{2;2}|^2+|\mathcal{P}_{2;3}|^2=1,\\
& \mathcal{P}_{1;1}\mathcal{P}^*_{2;1}+\mathcal{P}_{1;2}\mathcal{P}^*_{2;2} + \mathcal{P}_{1;3}\mathcal{P}^*_{2;3}=0.
\end{split}
\end{equation}
The success probability for this state is given by the norm of the output state,
\begin{equation}
\label{TwoQBProb}
	\mathrm{Pr}_{\rm cl}=4|\alpha_1\alpha_2t_1t_2|^2.
\end{equation}
This set of equations is a regular chain the solution of which can be obtained by solving the first equation for one variable, substituting the result into the second one, and then continuing until all the variables are expressed in terms of five variables, say $\mathcal{P}_{1;2},\alpha_{1,2}$ and $t_{1,2}$.
The step-by-step algorithm for numerical optimization is given in Appendix~\ref{AppC}.
The last step is to run an optimization for the probability of success~\eqref{TwoQBProb} over parameter ranges. 
We have depicted the probability of success of the scheme for a range of relative phase $0\leq\phi\leq\pi$ in Fig.~\ref{TwoQBClustPhi}.
In particular, the success probability at $\phi=\pi$ is $\mathrm{Pr}_{\rm cl}\approx 0.088$.
Once we have found the optimal solution for a particular phase $\phi$, we can use the well-known decomposition by Reck \etal~\cite{Reck} to evaluate the parameters of the linear optical network realizing $\hat{U}(3)$.
For the special case of $\chi=0$ and $\phi=\pi$, the resulting unitary matrix has been given in Appendix~\ref{AppD}.

We can compare the method used here against other schemes for achieving this particular output state.  Knill~\cite{Knill02} was motivated to find optimised probabilities for Controlled-Z rotations between optical qubits using linear interactions and post-selection in the Fock basis.  Knill's constructions were single-rail as for the case of the Controlled-Z rotations the single-rail operations are directly embedded in the dual-rail operations.  The result of Knill also did not have the multi-party preparation model that is used in this paper and hence considered arbitrary linear ineractions between all modes.  Knill found that when $\phi=\pi$ and with two ancillary modes containing single photons the highest probability of success was $2/27 \approx 0.074$.  Later extensive numerical searches confirmed this maximal probability over a wider range of networks with more ancillary modes~\cite{Uskov09}.  The ability to exceed this value for the device presented here can be attributed to the lack of need to construct a gate which operates over an entire Hilbert space.  The protocol here is merely the generating of the state and hence leaves open the possibility for probabilities exceeding that from implementations which utilise gates.

%============================================
\subsection{Two-qubit passive state transformation}

It is of interest to ask ``how the scheme of Fig.~\ref{TwoQBSch} may transform a given two-qubit input state?''
In the first place, our scheme is {\it passive} in the sense that there is no possibility for photon addition to any of the modes.
Secondly, one cannot switch photons between the two modes (bit flip) and this is exactly how we prevented the photon bunching effect in the output modes, i.e., the $|02\rangle$ and $|20\rangle$ components.
It is straight forward to calculate the output state similar to that of Eq.~\eqref{Out} starting with an arbitrary two-qubit state state
$|\psi_{\rm in}\rangle = a_{00}|00\rangle + a_{01}|01\rangle + a_{10}|10\rangle + a_{11}|11\rangle$.
It turns out that the coefficients of the output are just linear combinations of $d_{ij}$'s of Eqs.~\eqref{d00},~\eqref{3Eqs} and~\eqref{2Extra}.
A regular chain remains as a regular chain under linear combination of equations, and thus the resultant set always possesses a solution.
This means that our scheme is also a universal quantum state modifier for a known two-qubit inputs in a passive way.
Of course, starting with an input state within a subspace of $\mathcal{H}_1\otimes\mathcal{H}_2$ with only one or no photons means an overhead in the free parameters of the scheme.
However, such redundancy will be automatically resolved via optimization of success probability.

%============================================
\subsection{Effect of inefficiencies and losses}

Inefficiencies and losses are present in any experiment. 
Here, we analyse such deleterious effects, which are four fold: (i) imperfect single photon sources, (ii) the transmission losses through channels, (iii) losses through the unitary circuit, and (iv) inefficient measurements.
These imperfections directly affect the fidelity of the output state. 
Therefore, to simplify the analysis, we make the following assumptions:
\begin{enumerate}[(i)]
    \item All the measurement devices are the same, and thus they have equal losses,
    \item We model all the losses within the unitary network via symmetric losses on all input modes.
    \item All the photon sources are the same having equal purities, and,
    \item The losses within the principal output modes are equal to that of entangling measurement circuit.
\end{enumerate}
It is known that when the losses within all output ports of a linear network are equal, one can commute the loss with the network.
In this way, using the assumptions above, we may express all the imperfections just in terms of imperfect photon sources, and considering all the other elements to be ideal.
To justify the last assumption, we may think of a lossy measurement process on the output state such that it exactly equates the loss of  entangling measurement circuit.
We can therefore model all the losses using impure input states, i.e.,
\begin{equation}
    \hat{\varrho}_{\rm I}=(\mu |1\rangle\langle 1|+(1-\mu)|0\rangle\langle 0|)^{\otimes 2},
\end{equation}
in which $\mu$ is assumed to be the overall efficiency of the scheme.
Now, the un-normalized output, $\hat{\varrho}_{\rm AB}(\mu)$, is clearly a mixture of four input combinations, $|00\rangle_{12}$, $|10\rangle_{12}$, $|01\rangle_{12}$, and $|11\rangle_{12}$.
We have already calculated the output resulting from the last input, $|\psi_{\rm out}^{11}\rangle$.
The other three will give
\begin{align}
\label{ImpureTerms}
    &|\psi_{\rm out}^{00}\rangle=d_{00}|00\rangle_{\rm AB}=\frac{c_{11}\sqrt{\mathrm{Pr}_{\rm tar}}}{t_1t_2}|00\rangle_{\rm AB},\nonumber\\
    &|\psi_{\rm out}^{10}\rangle = t_1d_{00}|10\rangle_{\rm AB} + r_1d_{10}|00\rangle_{\rm AB}\nonumber\\
    &~~\qquad = \frac{\sqrt{\mathrm{Pr}_{\rm tar}}}{t_2}\Big(c_{11}|10\rangle_{\rm AB} + c_{01}|00\rangle_{\rm AB}\Big),\nonumber\\
    &|\psi_{\rm out}^{01}\rangle = t_2d_{00}|01\rangle_{\rm AB} + r_2d_{01}|00\rangle_{\rm AB}\nonumber\\
    &~~\qquad = \frac{\sqrt{\mathrm{Pr}_{\rm tar}}}{t_1}\Big(c_{11}|01\rangle_{\rm AB} + c_{10}|00\rangle_{\rm AB}\Big),
\end{align}
in which $\mathrm{Pr}_{\rm tar}$ is the probability of success for getting the desired target state when the input states are pure single photons.
Using the expressions in Eq.~\eqref{ImpureTerms} it is straight forward to evaluate the success probability as
\begin{align}
\label{Prmu}
    \mathrm{Pr}_{\rm tar}(\mu,|\psi_{\rm tar}\rangle) & = {\rm Tr}\hat{\varrho}_{\rm O}(\mu)\nonumber\\
    & = \mathrm{Pr}_{\rm tar}\Big[\mu^2  + (1-\mu)^2 A
     +\mu(1-\mu)B\Big],
\end{align}
in which
\begin{align}
    A=\frac{|c_{11}|^2}{t_1^2t_2^2},\quad B=\frac{|c_{11}|^2+|c_{01}|^2}{t_2^2}+\frac{|c_{11}|^2+|c_{10}|^2}{t_1^2}.
\end{align}
This result demonstrates the fact that, having imperfections, the heralding probability strongly depends on the target state as well as losses.
In a similar way, one can calculate the fidelity of the output state as
\begin{align}
\label{Fidmu}
    \mathfrak{F}(\mu,|\psi_{\rm tar}\rangle)&=\frac{\langle\psi_{\rm tar}|\hat{\varrho}_{\rm AB}(\mu)|\psi_{\rm tar}\rangle}{\mathrm{Pr}_{\rm tar}(\mu)}\nonumber\\
    &=\frac{\mu^2+(1-\mu)^2C + \mu(1-\mu)D}{\mu^2+(1-\mu)^2A + \mu(1-\mu)B},
\end{align}
in which
\begin{align}
    &C=|c_{00}c_{11}|^2,\nonumber\\
    &D=|c_{11}^\ast c_{10} + c_{01}^\ast c_{00}|^2 + |c_{11}^\ast c_{01}+c_{10}^\ast c_{00}|^2.
\end{align}
Expanding the fidelity around one, we find the rate at which it decays as
\begin{align}
    \frac{d}{d\mu}\mathfrak{F}(\mu,|\psi_{\rm tar}\rangle)\Big|_{\mu=1}= B-D,
\end{align}
or, equivalently, for nearly perfect efficiency ($\mu\rightarrow 1$)
\begin{equation}
\mathfrak{F}(\mu,|\psi_{\rm tar}\rangle)\approx1-(B-D)(1-\mu).
\end{equation}
For the example of a cluster state $D=0$ and thus, this rate is given by
\begin{align}
    \frac{d}{d\mu}\mathfrak{F}(\mu,|\psi_{\rm cl}\rangle)\Big|_{\mu=1}=\frac{1}{t_1^2}+\frac{1}{t_2^2}.
\end{align}

%==============================================================================
%==============================================================================
\section{Arbitrary three-qubit state preparation} \label{PO}

\begin{figure}[h]
  \includegraphics[width=8cm]{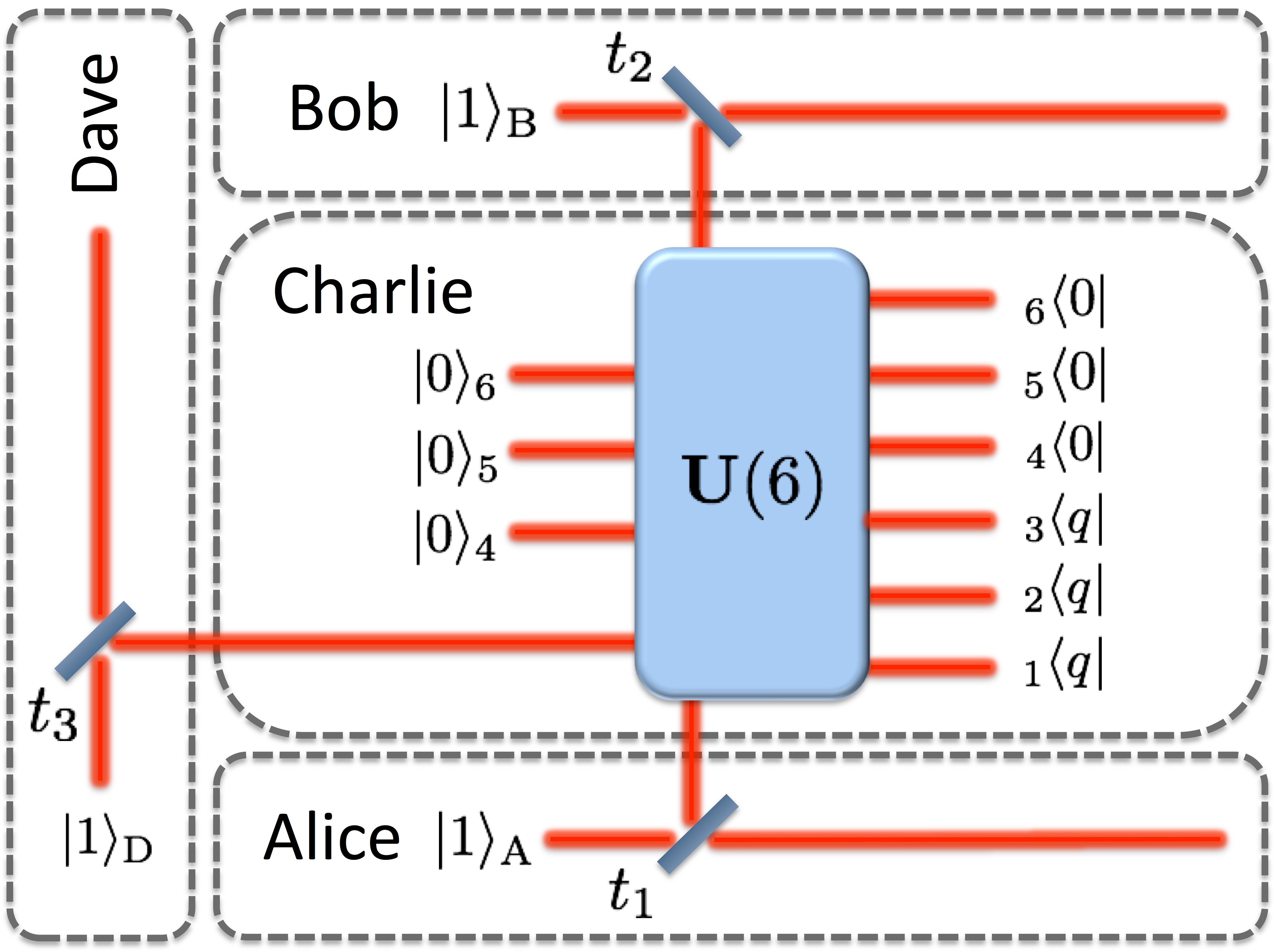}
  \caption{The scheme for universal three-qubit state generation.
  This time, Charlie uses a $\mathbf{U}(6)$ to perform the entangling measurement.
  }\label{ThreeQBSch}
\end{figure}

A similar strategy as the previous section can be taken to generate arbitrary three-qubit states.
We take three single-photon input states, split them and make an entangling measurement.
The scheme thus would consist of a $\mathbf{U}(6)$ matrix instead of $\mathbf{U}(3)$ as described in Appendix~\ref{AppB}, injection of three single photons and three ancillary vacua into the circuit, three arbitrary single-qubit and three vacuum measurements (c.f. Fig.~\ref{ThreeQBSch}).

After repeating the process, we find the projector
\begin{align}	
\label{phi123}
    _{123}\langle\varphi|&= {_{123}\langle}000|d_{000} +              {_{123}\langle}100|d_{100} +                         {_{123}\langle}010|d_{010}\nonumber\\
    & + {_{123}\langle}001| d_{001} + {_{123}\langle}110| d_{110} + {_{123}\langle}101|d_{101}\nonumber\\
    & + {_{123}\langle}011|d_{011} + {_{123}\langle}111|d_{111}\nonumber\\
    & + \text{terms containing two or three photons}\nonumber\\
    & \quad\text{in either of modes},
\end{align}
in which
\begin{align}
\label{3QBdijk}
    d_{000}&=\alpha_1\alpha_2\alpha_3\nonumber\\
	d_{100}&=\alpha_2\alpha_3\beta_1 \mathcal{P}_{1;1} +\alpha_1\alpha_3\beta_2 \mathcal{P}_{2;1} + \alpha_1\alpha_2\beta_3 \mathcal{P}_{3;1}\nonumber\\
	d_{010}& = \alpha_2\alpha_3\beta_1 \mathcal{P}_{1;2} +\alpha_1\alpha_3\beta_2 \mathcal{P}_{2;2} + \alpha_1\alpha_2\beta_3 \mathcal{P}_{3;2}\nonumber\\
	d_{001}&= \alpha_2\alpha_3\beta_1 \mathcal{P}_{1;3} +\alpha_1\alpha_3\beta_2 \mathcal{P}_{2;3} + \alpha_1\alpha_2\beta_3 \mathcal{P}_{3;3}\nonumber\\
	d_{110}&= \alpha_1\beta_2\beta_3 \mathcal{P}_{23;12} +\alpha_2\beta_1\beta_3 \mathcal{P}_{13;12} + \alpha_3\beta_1\beta_2 \mathcal{P}_{12;12}\nonumber\\
	d_{101}&= \alpha_1\beta_2\beta_3 \mathcal{P}_{23;13} +\alpha_2\beta_1\beta_3 \mathcal{P}_{13;13} + \alpha_3\beta_1\beta_2 \mathcal{P}_{12;13}\nonumber\\
	d_{011}&= \alpha_1\beta_2\beta_3 \mathcal{P}_{23;23} +\alpha_2\beta_1\beta_3 \mathcal{P}_{13;23} + \alpha_3\beta_1\beta_2 \mathcal{P}_{12;23}\nonumber\\
	d_{111}&= \beta_1\beta_2\beta_3\mathcal{P}_{123;123}.
\end{align}

As an example, we have determined the optimal probability of success for generation of a generalized Greenberger-Horne-Zeilinger (GHZ) state,
\begin{equation}
\label{GHZ}
 |\psi_{\rm GHZ}\rangle=\frac{1}{\sqrt{2}}(|000\rangle + e^{i\phi}|111\rangle),
\end{equation}
to be ${\rm Pr_{GHZ}}\approx 0.0024$.
Clearly, the probability of preparing such a state is independent of the phase value $\phi$, since it can be obtained by a local $\phi$-phase shift of one of the bits at the output.

In parallel with discussions of Sec.~\ref{SQProj}, we may represent events as $\{\Pi^{(i,j,k)}\equiv\Pi^{(i)}\otimes\Pi^{(j)}\otimes\Pi^{(k)}\}$ ($i,j,k=0,1,2$).
For this scheme as well, all the arguments of Sec.~\ref{SQProj} hold such that we may improve the heralding probability by a factor of 2, if we take into account not only the event $\Pi^{(1,1,1)}$, but also the invert event $\Pi^{(2,2,2)}$.
From Eqs.~\eqref{phi123} and~\eqref{3QBdijk}, it is evident that we may correct the output just by feed-forwarding three extra classical bits and operating locally on the principal output modes by phase shifters of angle $\pi$.

A second important thing to note is that, as we can see, all the permanents of the active submatrix of $\mathbf{U}(6)$ appear in the expressions for the coefficients of the projector. In the next section, we will show that this is generally true for arbitrary large number of qubits.
We also discuss the implication of this pattern about the scalability of our scheme, and we will see that this puts strong limitations on any measurement based scheme.

The non-deterministic remote state preparation of a single-rail GHZ state using linear optics and post-selection does not appear to have been proposed before.  
Though the generation probability here is low, to the best of authors' knowledge, there is no other schemes with which to fairly compare it to.

%==============================================================================
%==============================================================================
\section{Multi-qubit Scaling}

As we promised, in this section we show that our method can be generalized to prepare arbitrary $N$-qubit states.
Thus, from an architectural point of view, everything seem to be fine for having such a machine.
In Appedix~\ref{AppB}, we have discussed the size of the unitary matrix and the active submatrix to be considered.
In general, $M$ scales as $O(2^N)$, the same as the size of the active submatrix.
Let us point out that the active submatrix need not be square in general.
In fact, one can choose the size depending on the required number of free parameters by choosing any ancillary input port to be injected by a vacuum or a single photon state and any output port to be measured in a vacuum or a single-qubit basis. 
As a rule of thumb, the input-port number and the output-port number label the row and column index respectively.
For instance, if the output-port one is the only one being projected onto a single qubit, and input ports one and two are open to Alice and Bob whilst the rest are injected by vacuum state, then the active submatrix is
\begin{equation}
\label{S21}
	\mathbf{S}(2\times 1)=
	\begin{bmatrix}
	u_{11} \\ 
	u_{21} 
	\end{bmatrix}.
\end{equation}
If, in addition, Charlie decides to inject a single photon into the port three, then the resulting active submatrix will be
\begin{equation}
\label{S31}
	\mathbf{S}(3\times 1)=
	\begin{bmatrix}
	u_{11} \\
	u_{21} \\
	u_{31} 
	\end{bmatrix}.
\end{equation}

In the general scenario, we inject $N$ single-photons into $N$ principal modes, extracting a bit of each photon and processing them through a unitary of size $M(N)$.
The ports $1,2,\dots,N$ of $\mathbf{U}(M)$ are dedicated to principal inputs,
the ports $N+1,\dots,N+K$ are injected with ancillary single photons and the remaining $M-N-K$ with ancillary vacua.
At the output, single-qubit and vacuum measurements are performed on the ports $1,2,\dots,L$ and $L+1,\dots,M$, respectively.
Note that it is required to have $L\geqslant N+K$, otherwise there would be no $|1\rangle^{\otimes N}$ component in the output.
The projector $_{1\cdots N}\langle\varphi|$ in this case will have coefficients of the form
\begin{widetext}
\begin{align}
\label{NQBcoef}
    &d_{0_1\cdots 0_N}=\sum_{\sigma}\beta_{\sigma(1)}\cdots\beta_{\sigma(K)}\alpha_{\sigma(K+1)}\cdots\alpha_{\sigma(L)}\mathcal{P}_{\sigma(1)\cdots\sigma(K);(N+1)\cdots(N+K)},\nonumber\\
    &d_{1_10_2\cdots 0_N}=\sum_{\sigma}\beta_{\sigma(1)}\cdots\beta_{\sigma(K)}\beta_{\sigma(K+1)}\alpha_{\sigma(K+2)}\cdots\alpha_{\sigma(L)}\mathcal{P}_{\sigma(1)\cdots\sigma(K)\sigma(K+1);1(N+1)\cdots(N+K)},\nonumber\\
    &d_{0_11_2\cdots 0_N}=\sum_{\sigma}\beta_{\sigma(1)}\cdots\beta_{\sigma(K)}\beta_{\sigma(K+1)}\alpha_{\sigma(K+2)}\cdots\alpha_{\sigma(L)}\mathcal{P}_{\sigma(1)\cdots\sigma(K)\sigma(K+1);2(N+1)\cdots(N+K)},\nonumber\\
    &\qquad\vdots\nonumber\\
    &d_{1_11_2\cdots 0_N}=\sum_{\sigma}\beta_{\sigma(1)}\cdots\beta_{\sigma(K)}\beta_{\sigma(K+1)}\beta_{\sigma(K+2)}\alpha_{\sigma(K+3)}\cdots\alpha_{\sigma(L)}\mathcal{P}_{\sigma(1)\cdots\sigma(K)\sigma(K+1)\sigma(K+2);12(N+1)\cdots(N+K)},\nonumber\\
    &\qquad\vdots\nonumber\\
    &d_{1_11_2\cdots 1_N}=\sum_{\sigma}\beta_{\sigma(1)}\cdots\beta_{\sigma(N+K)}\alpha_{\sigma(N+K+1)}\cdots\alpha_{\sigma(L)}\mathcal{P}_{\sigma(1)\cdots\sigma(N+K);1\cdots(N+K)},
    %\mathcal{P}_{1\cdots N;1\cdots (N+K)}\prod_{i=1}^{N}\beta_i,
\end{align}
\end{widetext}
where the sums run over all the permutations $\sigma$ of $L$ index numbers.
Evidently, the size of the active submatrix is $L\times(N+K)$.

According to Eq~\eqref{NQBcoef}, to determine the elements of the active submatrix $\mathbf{S}$ and $\mathbf{U}(M)$ one should solve a set of polynomial equations of degree at most $N+K$ over complex numbers and optimize the probability of success.
The question is whether or not we can determine if the set of equations obtained by comparing Eq.~\eqref{NQBcoef} with the coefficients of any arbitrary target state possesses a solution and if so, can we find it?
This question can be answered for complex variables using the well-known Gr\"{o}bner basis and elimination techniques~\cite{DickensteinBOOK}.  
However, as noted in~\cite{VanMeter2007}, there is generally no efficient algorithm to calculate solutions to such sets of equation.  
Furthermore, we can note from Eq.~\eqref{NQBcoef} that merely writing down the problem to be solved is inefficient as in principle there may be exponentially many $d$ coefficients.  
If we posed a problem restricted to polynomially many $d$'s, then the sum over the entire symmetric group will be exponentially large.  
Finally, even if there were only polynomially many non-zero $\alpha$'s and $\beta$'s we are still left with evaluating matrix permanents, a problem known to be classically hard to compute, over polynomials of the unknown elements of the active submatrix $\mathbf{S}$.
As a consequence, we are encountering a \textit{verification problem}.
Therefore, the solution to the posed problem, although it might exist, cannot be efficiently found.
However, it is interesting to note that, in the case where there is polynomially many $d$ coefficients and a polynomial error bound, a universal quantum computer would very likely be able to implement the projector $_{1\cdots N}\langle\varphi|$ on qubits efficiently.  

The other important question is how does the success probability scale?
Unfortunately there is no unique answer to this question, because the success probability strongly depends on the target state.
For example, preparing $N$-mode separable single-photon or vacuum state is trivially possible with success probability of $1$ irrespective of the number of modes or measurements.
Similarly, assuming that we could perform deterministic single-qubit measurements at the output of $\mathbf{U}(M)$, by making use of a feed-forward strategy and correcting the phase at the output, the probability of success for preparing any $N$-mode separable state would be equal to one.
Including the success probability of the single-qubit measurements, however, causes the scale of the scheme to be $O(1/p^N)$, where $p$ is the success probability of one single-qubit measurement, even for generating separable states.

%==============================================================================
%==============================================================================
\section{Outlook and Conclusion}

In conclusion, we have proposed a technique for the generation of arbitrary multi-qubit quantum states using linear optics and heralding measurements in a single-rail encoding scenario.
Our approach avoids any multi-photon terms at the output. 
We explicitly showed that our scheme leads to the set of equations, involving permanents of a submatrix of the linear optical network, which is solvable for a small number of qubits.
In particular, we solved the problem for generalized cluster states of two- and three-qubits as target states and obtained a heralding probability for our universal scheme which is comparable to the state-of-the-art proposals for specific states alone.
We also considered the scalability of our scheme and discussed the efficiency of finding a solution to the general set of equations from various perspectives.
It turned out that a universal quantum computer is the only hope for solving the general problem.
Our proposal thus can be reliably used for sharing universal multi-qubit quantum states between a few number of parties.
Moreover, it can be applied equally to other quantum optical systems such as opto-mechanics and spin ensembles.

%==============================================================================
%==============================================================================
\section*{Acknowledgements}

The authors acknowledge useful discussions by S. Rahimi-Keshari and C. Schaeff.
This project was supported by the Australian Research Council Discovery Project (No. DP140101638) and the Australian
Research Council Centre of Excellence of Quantum Computation and Communication Technology (Project No. CE110001027).

%==============================================================================
%==============================================================================
%==============================================================================
%==============================================================================
\appendix

\section{Existence of a solution to Eq.~\eqref{Set1}} 
\label{AppA}

To prove that we have enough control to produce arbitrary two-qubit states, we need to prove that the set of Eq.~\eqref{Set1} forms a regular chain.
We have
\begin{align}
\label{CH1}
\begin{split}
&f_1:\quad \frac{r_1r_2\beta_1\beta_2(u_{11}u_{22}+u_{12}u_{21})}{p}-c_{00}=0,\\
&f_2:\quad \frac{r_1t_2(\alpha_1\beta_2u_{21}+\alpha_2\beta_1u_{11})}{p}-c_{01}=0,\\
&f_3:\quad \frac{r_2t_1(\alpha_1\beta_2u_{22}+\alpha_2\beta_1u_{12})}{p}-c_{10}=0.
\end{split}
\end{align}
They involve three independent variables from the set $\{u_{11},u_{12},u_{22},u_{21}\}$.
Suppose that we choose $u_{11}$, $u_{21}$, and $u_{22}$ as free.
Now, it is important to note that $f_2$ and $f_3$ are independent equations.
Considering $f_1$ and $f_2$, they have the variables $u_{11}$ and $u_{21}$ in common.
They have a common solution over the field of complex numbers if and only if they possess a vanishing \emph{resultant}~\cite{DickensteinBOOK}.
One common procedure to find the solution is to first evaluate the resultant with respect to one variable,
\begin{equation}
\begin{split}
{\rm Res}_{1,1}(f_1,f_2;u_{11})&=|{\rm Syl}(f_1,f_2;u_{11})|\\
&=\begin{vmatrix}
	S_{00}u_{22} & S_{00}u_{12}u_{21}-c_{00}\\ 
	S_{10} & S_{11}u_{21}-c_{01}
	\end{vmatrix}\\
&=0,
\end{split}
\end{equation}
where ${\rm Syl}(Q,P,x)$ is the Sylvester matrix of the polynomials $Q$ and $P$ with respect to the variable $x$,
$S_{00}=r_1r_2\beta_1\beta_2/p$, $S_{10}=r_1t_2\alpha_2\beta_1/p$, and $S_{11}=r_1t_2\alpha_1\beta_2/p$.
This gives the \emph{unique} solution for $u_{21}$ and putting it back into $f_2$ gives the solution for $u_{11}$.
A simple calculation then leads to
\begin{equation}
\begin{split}
&u_{21}=\frac{S_{00}c_{01}u_{22}-S_{10}c_{00}}{S_{00}(S_{11}u_{22}-S_{10}u_{12})},\\
&u_{11}= \frac{c_{01}}{S_{10}}-\left(\frac{S_{11}}{S_{10}}\right)\left(\frac{S_{00}c_{01}u_{22}-S_{10}c_{00}}{S_{00}(S_{11}u_{22}-S_{10}u_{12})}\right).
\end{split}
\end{equation}
Now, we have both solutions in terms of the remaining free variable, $u_{22}$.
Note that $u_{12}$ is determined from conditions on $\mathbf{U}(3)$.
Therefore, $f_3$ also uniquely determines $u_{22}$.
This analysis completes the proof that $\mathbf{U}(3)$ is the smallest unitary with enough degrees of freedom for arbitrary entangling measurements in our scheme.

Notably to say, there is also a second method to prove that the set~\eqref{CH1} has a solution, and that is to show that the Macaulay's resultant of the set is zero (see Ref.~\cite{Jonsson} for a detailed account of the method).

%==============================================================================
%==============================================================================
\section{Size of the optimal unitary}
\label{AppB}

Let us consider the general case of $N$ qubits.
An arbitrary state has $2^N-1$ free complex coefficients.
Therefore, we need a unitary of the size $M$ such that it has enough free magnitude and phase degrees of freedom.
Any $M\times M$ unitary has $M(M+1)/2$ free parameters, $M(M-1)/2$ of which are magnitudes and $M$ of them phases.
One should also notice that one of the phases is global which makes the difference between $\mathbf{U}(M)$ and $\mathbf{SU}(M)$ (the $M\times M$ special unitary matrices).
% in the same way as discussed in Sec.~\ref{PO}.
Therefore, $M$ must satisfy
\begin{equation}
\frac{M(M-1)}{2}\geqslant 2^N-1\quad\text{and}\quad
M\geqslant 2^N-1.
\end{equation}
We note that
% except for $N=1$, 
the second condition has the a slower growth in $M$, and thus determines $M$ as
\begin{equation}
M=2^N-1\quad\text{for}\quad N\geqslant 2.
\end{equation}
For $N=1$ there is no need for a unitary at all, since there is enough degrees of freedom in measurement and pick-off.

Besides this general consideration, $M$ could be chosen to be smaller or larger based on the particular value of $N$.
This is because for $M$ to have enough free parameters is necessary, but the more important is to chose the active submatrix of the right dimensions to involve enough free parameters in the equations.
For example, in the case of $N=3$ of Sec.~\ref{PO}, the coefficient $c_{111}\propto d_{000}$ is uniquely determined via the measurements; cf. Eq.~\eqref{3QBdijk}.
Thus, the global phase of the unitary can effectively cause a relative phase between $|111\rangle$ term and the rest.
As a result, instead of a $\mathbf{U}(7)$, a $\mathbf{U}(6)$ delivers enough free parameters when a $3\times 3$ active submatrix is chosen.

%==============================================================================
%==============================================================================
\section{Step-by-step algorithm for solving Eqs.~\eqref{setU} and~\eqref{setConst}}
\label{AppC}

To solve Eqs.~\eqref{setU} and~\eqref{setConst} numerically, we follow the procedure:
\begin{enumerate}[(1)]
\item Solve the first equation in~\eqref{setU} for $\mathcal{P}_{2;2}$ in terms of $\mathcal{P}_{1;2}$.
\item Solve the second equation in~\eqref{setU} for $\mathcal{P}_{2;1}$ in terms of $\mathcal{P}_{1;1}$.
\item Substitute the results into the third equation and solve it for $\mathcal{P}_{1;1}$ to obtain it as a function of $\mathcal{P}_{1;2}$.
\item Substitute the result back into the second equation to transform $\mathcal{P}_{2;1}$ in terms of $\mathcal{P}_{1;2}$.
At this stage, we have all the parameters as functions of variables $\mathcal{P}_{1;2},\alpha_{1,2},t_{1,2}$, and $\zeta$.
\item Notice that $\mathcal{P}_{1;3}$ is directly given in terms of $\mathcal{P}_{1;2}$ from the first constraint in~\eqref{setConst}, since $\mathcal{P}_{1;1}$ is obtained in step (3).
\item Now, we are left with two unknowns, $\mathcal{P}_{1;2}$ and $\mathcal{P}_{2;3}$.
They can be obtained from the second and third constraints in~\eqref{setConst}. 
However, we prefer to calculate $\mathcal{P}_{2;3}$ from the last one in terms of $\mathcal{P}_{1;2}$.
\item As the last step, we run an optimization of the success probability, Eq.~\eqref{TwoQBProb}, over variables $\mathcal{P}_{1;2},\alpha_{1,2},t_{1,2}$, and $\zeta$ constrained to the second condition in Eq.~\eqref{setConst}.
\end{enumerate}

%==============================================================================
%==============================================================================
\section{Unitary elements for a cluster state}
\label{AppD}
Here we give our numerical result for the $\mathbf{U}(3)$ leading the highest probability of success for generating a two-qubit cluster state (see Eq.~\eqref{TwoCluster}) of $\chi=0$ and $\phi=\pi$.
\begin{widetext}
\begin{equation}
\mathbf{U}_{\rm cl}(3)=
\left(
\begin{array}{ccc}
 -0.493-0.312 i & -0.493-0.312 i & 0.565 \\
 0.338\, +0.214 i & 0.338\, +0.214 i & 0.825 \\
 0.593\, -0.384 i & -0.593\, +0.384 i & 0 \\
\end{array}
\right).
\end{equation}
The corresponding optimal values for other parameters are
\begin{equation}
\begin{split}
&\alpha_1=0.452,\\
&\alpha_2=0.791,\\
&t_1=t_2=0.645,\\
&\zeta = 2.577.
\end{split}
\end{equation}
\end{widetext}


\begin{thebibliography}{100}




\bibitem{Nielsen} Nielsen M A, and Chuang I L 2000 \textit{Quantum Computation and Quantum Information}, (Cambridge University Press)

\bibitem{Kimble2008} Kimble H J \Nature{453}{1023}{2008}{The quantum internet}

\bibitem{Guzik2012} Aspuru-Guzik A, and Walther P \NatPhys{8}{285}{2012}{Photonic quantum simulators}

\bibitem{Giovannetti2011} Giovannetti V, Lloyd S, and Maccone L \NatPhot{5}{222}{2011}{Advances in quantum metrology}

\bibitem{Leibfried2003} Leibfried D, Blatt R, Monroe C, and Wineland D \RMP{75}{281}{2003}{Quantum dynamics of single trapped ions}

\bibitem{Hammerer2010} Hammerer K, Sorensen A S, and Polzik E S \RMP{82}{1041}{2010}{Quantum interface between light and atomic ensembles}

\bibitem{Doherty2013} Doherty M W, Manson N B, Delaney P, Jelezko F, Wrachtrup J, Hollenberg L C L \PhysRep{1}{528}{2013}{The nitrogen-vacancy colour centre in diamond}

\bibitem{Glockl2003} Gl{\"{o}}ckl O \etal~\PRA{68}{012319}{2003}{Experiment towards continuous-variable entanglement swapping: Highly correlated four-partite quantum state}

\bibitem{Patel2016} Patel R B \etal~\SA{2}{e1501531}{2016}{A quantum Fredkin gate}

\bibitem{Browne2005} Browne D E and Rudolph T, \PRL{95}{010501}{2005}{Resource-Efficient Linear Optical Quantum Computation}
\bibitem{Lu2007} Lu C-Y~\etal~\NatPhys{3}{91}{2007}{Experimental entanglement of six photons in graph states}

\bibitem{Carolan2015} Carolan J \etal~\Science{349}{711}{2015}{Universal linear optics}
\bibitem{Schaeff2015} Schaeff C, Polster R, Huber M, Ramelow S, and Zeilinger A \Optica{2}{523}{2015}{Experimental access to higher-dimensional entangled quantum systems using integrated optics}


\bibitem{Dakna1999}Dakna M, Clausen J, Kn{\"{o}}ll L, and Welsch D.-G \PRA{59}{1658}{1999}{Generation of arbitrary quantum states of traveling fields}

\bibitem{Yukawa2013} Yukawa M \etal~\OEx{21}{5529}{2013}{Generating superposition of up-to three photons for continuous variable quantum information processing}

\bibitem{Brannan2014} Brannan T, Qin Z, MacRae A, and Lvovsky A I \OL{39}{5447}{2014}{Generation and tomography of arbitrary qubit states using transient collective atomic excitations}



\bibitem{VanMeter2007} VanMeter N M \etal~\PRA{76}{063808}{2007}{General linear-optical quantum state generation scheme: Applications to maximally path-entangled states}
\bibitem{Andersen2015} Andersen U L, Neergaard-Nielsen J S, van Loock P, and Furusawa A~\NatPhys{713}{11}{2015}{Hybrid discrete- and continuous-variable quantum information}


\bibitem{Lund2002} Lund A P and Ralph T C~\etal~\PRA{66}{032307}{2002}{Nondeterministic gates for photonic single-rail quantum logic}

\bibitem{Simon2007} Simon C~\etal~\PRL{98}{190503}{2007}{Quantum Repeaters with Photon Pair Sources and Multimode Memories}

\bibitem{AA} Aaronson S and Arkhipov A \TC{9}{143}{2013}{The Computational Complexity of Linear Optics}.

\bibitem{Vanner2013} Vanner M R, Aspelmeyer M, and Kim M S \PRL{110}{010504}{2013}{Quantum State Orthogonalization and a Toolset for Quantum Optomechanical Phonon Control}

\bibitem{McConnell2015} McConnell R, Zhang H, Hu J, Cuk S, and Velutic V \Nature{519}{439}{2015}{Entanglement with negative Wigner function of almost 3,000 atoms heralded by one photon}

\bibitem{Wu} Wu L-A, Walther P, and Lidar D A \SR{3}{1394}{2013}{No-go theorem for passive single-rail linear optical quantum computing}

\bibitem{Pegg} Pegg D, Phillips L, and Barnett S \PRL{81}{1604}{1998}{Optical State Truncation by Projection Synthesis}


\bibitem{Lund} Ralph T C, Lund A P, and Wiseman H M \JOB{7}{S245}{2005}{Adaptive phase measurements in linear optical quantum computation}

\bibitem{Wiseman98} Wiseman H M and Killip R B \PRA{57}{2169}{1998}{Adaptive single-shot phase measurements: The full quantum theory}

\bibitem{Reck} Reck M, Zeilinger A, Bernstein H J, and Bertani P \PRL{73}{58}{1994}{Experimental Realization of Any Discrete Unitary Operator}

\bibitem{Knill02} Knill E \PRA{66}{052306}{2002}{Quantum gates using linear optics and postselection}

\bibitem{Uskov09} Uskov D B, Kaplan L, Smith A M, Huver S D, and Dowling J P \PRA{79}{042326}{2009}{Maximal success probabilities of linear-optical quantum gates}

\bibitem{DickensteinBOOK} Dickenstein A and Emiris I Z  2005 \textit{Solving Polynomial
Equations: Foundations, Algorithms, and Applications}, (Springer-Verlag Berlin Heidelberg)

\bibitem{Jonsson} J\'{o}nsson G F and Vavasis S A \MC{74}{221}{2004}{Accurate solution of polynomial equations using Macaulay resultant matrices}


\end{thebibliography}
\end{document}